\documentclass[aps,prb,amsmath,twocolumn]{revtex4}

\usepackage{bm}
\usepackage{graphicx}

\begin{document}

\title{Shot noise and Coulomb effects on non-local electron transport in normal-superconducting-normal heterostructures}
\author{Dmitri S. Golubev$^1$ and Andrei D.
Zaikin$^{1,2}$} \affiliation{$^1$ Institut f\"ur Nanotechnologie,
Karlsruher Institut f\"ur Technologie (KIT), 76021 Karlsruhe, Germany}
\affiliation{$^2$I.E. Tamm Department of
Theoretical Physics, P.N. Lebedev Physics Institute, 119991
Moscow, Russia}

\begin{abstract}
We argue that Coulomb interaction can strongly influence non-local electron transport in normal-superconducting-normal structures and emphasize direct relation between Coulomb effects and non-local shot noise. In the tunneling limit non-local differential conductance is found to have an S-like shape and can turn negative
at non-zero bias. At high transmissions crossed Andreev reflection yields positive noise cross-correlations and Coulomb anti-blockade of non-local electron transport.
\end{abstract}


\maketitle

\section{Introduction}

Discreteness of electron charge has a number of fundamental physical consequences, such as, e.g., shot noise in mesoscopic
conductors \cite{BB} and Coulomb blockade of charge transfer in tunnel junctions \cite{SZ}. About 10 years ago it was realized \cite{GZ01,LY1}
that these two seemingly different phenomena are closely related to each other: Coulomb blockade turns out to be stronger in conductors
with bigger shot noise. This fundamental relation was subsequently confirmed in experiments \cite{Pierre}. A close link between shot noise and Coulomb blockade exists not only in normal conductors but also in hybrid normal-superconducting ($NS$) structures \cite{GZ09}, where doubling of elementary charge due to Andreev reflection becomes important.

Can the above relation be further extended to include non-local effects? A non-trivial example is provided by normal-superconducting-normal ($NSN$) systems where entanglement between electrons in different normal terminals can be realized. Non-local electron transport in such systems is determined by an interplay between elastic cotunneling (EC) and crossed Andreev reflection (CAR) and was recently investigated both experimentally \cite{Beckmann,Teun,Venkat,Basel,Beckmann2} and theoretically \cite{FFH,KZ06,GKZ} (see also further refs. therein).
While non-interacting theory predicts that CAR never dominates over direct electron transfer (hence, no sign change of non-local signal could occur), both positive and negative non-local signals have been detected \cite{Beckmann,Teun,Basel,Beckmann2}. It was argued that CAR could prevail over EC in the presence of
Coulomb interactions \cite{LY} or an external ac field \cite{GlZ09}. Negative non-local conductance
was also predicted in interacting single-level quantum dots in-between normal and superconducting terminals \cite{Koenig}.

Despite these developments no general theory describing the effect of electron-electron interactions on non-local transport in $NSN$ structures was available until now. Below we will construct such a theory and demonstrate that interaction effects in non-local transport and non-local shot noise in such systems are intimately related. This relation, however, turns out to be much more subtle than in the local case \cite{GZ01,LY1,GZ09} merely because of ($a$) a variety of different processes contributing to non-local shot noise and ($b$) positive cross-correlations which may occur in normal-superconducting hybrids  \cite{BB,AD} (in contrast to normal conductors where cross-correlations of fluctuating currents are negative \cite{BB}). In tunnel $NSN$ systems EC and CAR provide respectively negative and positive contributions to non-local shot noise \cite{Pistolesi,Chandrasekhar}. Here we will analyze non-local shot noise beyond the tunneling limit and find
that at higher transmissions also direct electron transfer can yield positive cross-correlations {\it in addition} to CAR. At full transmissions only positive cross-correlations due to CAR survive and yield {\it Coulomb anti-blockade} of non-local electron transport.

The paper is organized as follows. In Sec. 2 we describe our model and derive an effective action for $NSN$ system under consideration. In Sec. 3 we formulate the Langevin equations describing real time dynamics of fluctuating voltages and currents and derive the general expressions for both local and non-local current-current correlators describing shot noise in our system at arbitrary barrier transmissions and arbitrary frequencies. Sec. 4 is devoted to the effects of Coulomb interaction on both local and non-local conductances of our $NSN$ device. A brief summary of our key observations is presented in Sec. 5. Some technical details are outlined in Appendix.

\begin{figure}
\includegraphics[width=5cm]{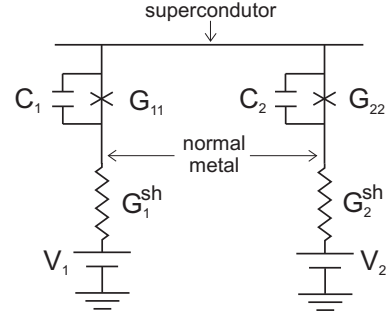}
\caption{Schematics of the system under consideration.}
\label{shema}
\end{figure}

\section{The model and effective action}
We will consider a hybrid structure consisting of two normal electrodes coupled to a superconductor via $NS$ barriers with local
subgap conductances $G_{11}$ and $G_{22}$ and capacitances $C_1$ and
$C_2$ (Fig. 1). External voltages $V_1$ and $V_2$ are applied to normal electrodes via
Ohmic shunts with conductances $G_{1}^{\rm sh}$ and $G_{2}^{\rm sh}$.  Weak electromagnetic coupling between two
$NS$ barriers (e.g. via modes propagating in the superconductor \cite{LY}) will be disregarded.
The Hamiltonian of the system reads
\begin{eqnarray}
H=H_1+H_2+H_S+H_{T,1}+H_{T,2},
\end{eqnarray}
where
%\begin{eqnarray}
$$
H_r=\sum_{\alpha=\uparrow,\downarrow}\int d{\bm x}
\hat \psi^\dagger_{r,\alpha}\left(-\frac{\nabla^2}{2m}-\mu\right)\hat \psi_{r,\alpha}, \;\;r=1,2,
$$
%\end{eqnarray}
are the Hamiltonians of the normal metals, $m$ is electron mass, $\mu$ is the chemical potential,
%\begin{eqnarray}
$$
H_S = \int d{\bm x} \bigg[
\sum_{\alpha}\hat\chi^\dagger_{\alpha}\left(-\frac{\nabla^2}{2m}-\mu\right)\hat\chi_{\alpha}
+\Delta \hat\chi^\dagger_{\uparrow}\hat\chi^\dagger_{\downarrow}
+ \Delta^* \hat\chi_{\downarrow}\hat\chi_{\uparrow}
\bigg]
$$
%\end{eqnarray}
is the Hamiltonian of the superconductor with order parameter $\Delta$ and
\begin{eqnarray}
H_{T,r}={\cal A}_r\sum_{\alpha,\beta=\uparrow,\downarrow}
\big[ t_r\, e^{i\varphi_r}\,  \hat\psi^\dagger_{\beta} \hat\chi_{r,\alpha}
+ t_r^*\, e^{-i\varphi_r}\,  \hat\chi_{r,\alpha}^\dagger\hat\psi_{\beta}  \big]
\label{HT}
\end{eqnarray}
are tunneling Hamiltonians describing transfer of electrons across the contacts with area ${\cal A}_r$ and tunneling amplitude $t_r$. For the sake of simplicity we will assume that both $NS$ barriers are uniform
implying that all $N_r=k_F^2{\cal A}_r/4\pi$ conducting channels in the $r$-th barrier are characterized by equal transmission values
\begin{equation}
T_{r}={4\pi^2\nu_r\nu_S|t_{r}|^2}/{\left(1+\pi^2\nu_r\nu_S|t_{r}|^2\right)^2}.
\label{ttr}
\end{equation}
where $\nu_j$ ($j=1,2,S$) is the density of states in the corresponding terminal. Accordingly, in the low energy limit to be considered below local subgap conductances are defined as \cite{BTK} $G_{rr}=(2e^2/\pi)N_r\tau_r$, where $\tau_r = T_r^2 / (2-T_r)^2$ are effective Andreev transmissions of $NS$ barriers.
Finally, we note that fluctuating phases $\varphi_r$ introduced in Eq. (\ref{HT})
are linked to the voltage drops across the barriers $v_r$ by means of the standard relation $\dot\varphi_r = ev_r$ and are treated as quantum operators.

As usually, we eliminate fermionic variables
and express the kernel $J$ of the Keldysh evolution operator
via path integral over the phase fields \cite{SZ}
\begin{equation}
J=\int \prod_{r=1,2}{\cal D} \varphi^F_r{\cal D}\varphi_r^B
\exp (iS_{\rm env}[\varphi ]+iS_{T}[\varphi ]), \label{pathint}
\end{equation}
where $\varphi^F_r$ and $\varphi_r^B$ are fluctuating phases defined respectively on the forward and backward branches of the Keldysh contour, $S_{\rm env}$ is the action of electromagnetic environment and the term $iS_{T}$ accounts for electron transfer between the terminals. In the case of linear Ohmic environment
considered here one has \cite{SZ}
\begin{eqnarray}
iS_{\rm env} &=& \sum_{r=1,2}\bigg[i\int dt \frac{(eV_{r}-\dot\varphi_r)(-C_r\dot\varphi_r^- +G_r^{\rm sh}\varphi^-_r)}{e^2}
\nonumber\\ &&
-\,\frac{G_r^{\rm sh}}{2e^2}\int dt dt'\varphi^-_r(t)M(t-t')\varphi^-_r(t')\bigg],
\end{eqnarray}
where $\varphi_r=(\varphi_r^F+\varphi_r^B)/2$, $\varphi^-_r=\varphi_r^F-\varphi^B_r$
and 
$$
M(t)=\int\frac{d\omega}{2\pi} e^{i\omega t}\omega\coth\frac{\omega}{2T}=-\frac{\pi T^2}{\sinh^2(\pi Tt)}.
$$
The term $iS_{T}$ reads
\begin{eqnarray}
iS_{T}={\rm tr}\ln  {\cal G}^{-1},\quad {\cal G}^{-1}=\left(
\begin{array}{ccc}
\check G_1^{-1} & \check t_1 & 0 \\
\check t_1^\dagger & \check G_S^{-1} & \check t_2 \\
0 & \check t_2^\dagger & \check G_2^{-1}
\end{array}\right),
\end{eqnarray}
where $4\times 4$ matrices $\check G_{j}^{-1}$ represent the inverse Keldysh Green functions 
of (isolated) normal ($j=1,2$) and superconducting ($j=S$) terminals and $\check t_r$ is diagonal $4\times 4$ matrix
in the Nambu - Keldysh space
\begin{eqnarray}
\check t_r = \left(
\begin{array}{cccc}
-t_r e^{-i\varphi_r^F} & 0 & 0 & 0 \\
0 & t_r e^{-i\varphi_r^B} & 0 & 0 \\
0 & 0 & t_r e^{i\varphi_r^F} & 0 \\
0 & 0 & 0 & -t_r e^{i\varphi_r^B}
\end{array}
\right).
\label{t}
\end{eqnarray}

After some exact manipulations we obtain
\begin{eqnarray}
iS_{T} =
\,{\rm tr}\,\ln\big[ \check 1 - \check t_1^\dagger \check G_1\check t_1\check G_S
- \check t_2^\dagger \check G_2\check t_2 \check G_S \big].
\label{S}
\end{eqnarray}
While the expression (\ref{S}) for the action remains formally exact it is still too complicated to be directly employed in our calculations. In order to proceed we will make several additional steps which yield necessary simplifications.

As a first step, we restrict ourselves to the limit of high conductances
\begin{equation}
g_r=2\pi(G_r^{\rm sh}+G_{rr})/e^2 \gg 1,
\end{equation}
in which case phase fluctuations are weak and
it suffices to expand the action (\ref{S}) to the second order in $\varphi^-_r$, cf., e.g., \cite{GZ01,GZ09,Schmid}.
Technically, we first expand the matrices $\check t_r$ to the second order in the quantum phases $\varphi^-_r$, i.e. we make a replacement
\begin{eqnarray}
\check t_r \to \check t_r(\varphi_r) \left(\check 1 - i\frac{\varphi^-_r}{2}\check\Lambda -\frac{\varphi_r^{-2}}{4}\check 1\right).
\end{eqnarray}
Here $\check t_r(\varphi_r)$ is defined by Eq. (\ref{t}) with $\varphi^{F,B}_r$ being replaced by the classical phase $\varphi_r$,
and $\check\Lambda$ is a diagonal matrix with non-zero elements $\Lambda_{11}=1$, $\Lambda_{22}=-1$, $\Lambda_{33}=-1$ and $\Lambda_{44}=1$. Accordingly, we can write the product $\check t_r^\dagger \check G_r\check t_r$ in the form
\begin{eqnarray}
\check t_r^\dagger \check G_r\check t_r =
\check\Sigma_r + \check\Sigma_r^{(1)} +  \check\Sigma_r^{(2)}+{\cal O}(\varphi^{-3}_r),
\end{eqnarray}
where we defined the self-energies
\begin{eqnarray}
\check\Sigma_r &=& \check t_r^{\dagger}(\varphi_r) \check G_r\check t_r(\varphi_r),
\label{sigma0}
\\
\check\Sigma_r^{(1)} &=& \frac{i}{2}\big[ \varphi^-_r\check\Lambda,\check\Sigma_r\big],
\nonumber\\
\check\Sigma_r^{(2)} &=& -\frac{1}{4}\big\{ \varphi^{-2}_r\check 1,\check\Sigma_r\big\}
+\frac{1}{4} \varphi^-_r\check\Lambda \check\Sigma_r \varphi^-_r\check\Lambda.
\nonumber
\end{eqnarray}
In order to evaluate $\check\Sigma_r$
we employ the Keldysh Green functions of the normal leads
\begin{eqnarray}
\check G_r=(\hat G_r^R+\hat G^A_r)\otimes \hat\sigma_z/2+(\hat G_r^R-\hat G^A_r)
\otimes \hat Q_r\hat\sigma_z/2.
\label{Gnormal}
\end{eqnarray}
Here the $2\times 2$ matrices $\hat G_r^{R,A}$ and $\hat Q_r$ are defined as
\begin{eqnarray}
\hat G_r^{R,A}=\left(\begin{array}{cc}
G^{R,A}_r & 0 \\
0 & G^{R,A+}_r
\end{array}\right), \; \hat Q_r=\left( \begin{array}{cc} 1-2n_r & 2n_r \\ 2-2n_r & 2n_r-1 \end{array} \right),
\nonumber
\end{eqnarray}
where $n_r$ is the quasiparticle distribution function in the $r$-th normal lead. In equilibrium it coincides with the Fermi function $n_r=1/(1+\exp(E/T))$.
Neglecting the proximity effect in the normal leads and
performing the summation over the corresponding electron states,
we express the zeroth order self-energies (\ref{sigma0}) in the form
\begin{eqnarray}
\check\Sigma_r(\varphi_r,\bm{r}) = \frac{\Gamma_r}{2i} h_r(\bm{r})
\left(\begin{array}{c}
\hat\sigma_z e^{i\varphi_r}\hat Q_r e^{-i\varphi_r} \;\;\;\;\;\; 0 \\
0 \;\;\;\;\;\; \hat\sigma_z e^{-i\varphi_r}\hat Q_r e^{i\varphi_r}
\end{array}\right),
\end{eqnarray}
with $\Gamma_r=2\pi\nu_r|t_r|^2$.
The function $h_r(\bm{r})$ in this expression differs from zero only at the interface of the $r$-th junction and it obeys the following normalization condition
\begin{equation}
\int d^3\bm{r}\,h_r(\bm{r})=N_r.
\label{norc}
\end{equation}
For the sake of simplicity, in what follows we will assume that
the barrier cross-sections remain sufficiently small and put
$h_r(\bm{r})=N_r\delta(\bm{r}-\bm{r}_r)$. This assumption just implies that each of the barriers has $N_r$ conducting channels with identical transmissions $T_r$ (\ref{ttr}, as we already indicated above. In this case we can reduce the full coordinate
dependence of the Green functions to that on the two indices $i$ and $j$ which label the barriers and, hence, can take only two values 1 and 2. Accordingly, e.g., the Green function $\check G_S(\bm{r},\bm{r}')$ reduces to the $2\times 2$ matrix in the "junction space" $\check G_S^{ij}$. In addition we should bear in mind that $\check G_S^{ij}$ are the matrices in the space of conducting channels.

We note that the above assumptions are not really restrictive since they do not affect the general
structure of our effective action to be derived below.
At the same time they allow to establish relatively simple expressions for the parameters entering in the action.
Expanding the action (\ref{S}) in powers of $\varphi^-_r$ we arrive at the following expression
\begin{eqnarray}
iS_T &=& -\,{\rm tr}\,\left[ \check K\left(\check\Sigma_1^{(1)}+\check\Sigma_2^{(1)}\right)\check G_S \right]
\nonumber\\ &&
-\,{\rm tr}\,\left[ \check K\left(\check\Sigma_1^{(2)}+\check\Sigma_2^{(2)}\right)\check G_S \right]
\nonumber\\ &&
-\,\frac{1}{2}\,{\rm tr}\,\left[ \left(\check K\left(\check\Sigma_1^{(1)}+\check\Sigma_2^{(1)}\right)\check G_S\right)^2  \right],
\label{S2}
\end{eqnarray}
where we define the operator
\begin{eqnarray}
\check K = \big[ 1 - \check\Sigma_1\check G_S- \check\Sigma_2 \check G_S \big]^{-1}.
\label{K1}
\end{eqnarray}

Our second step allows  to establish an explicit expression for the operator (\ref{K1}). Namely, in the interesting for us low energy limit
\begin{equation}
T, eV_r \ll |\Delta |
\label{le}
\end{equation}
we can set the energy argument $E$ in the superconductor Green function $\check G_S(E)$ equal to zero. After that
$\check G_S$ reduces to the time/energy independent matrix
\begin{eqnarray}
\check G_S(E=0)=\big( G_S^R+ G_S^A\big)\otimes ({\hat\sigma_z}/{2}),
\label{GS}
\end{eqnarray}
where we introduced the
retarded and advanced Green functions of the superconductor $G^R_S$ and $G^A_S$.
In the limit $E\to 0$ they are equal to each other both being $4\times 4$ matrices in the Nambu $\otimes$ "junction" space
\begin{eqnarray}
 G_S^R =G_S^A =
\left(\begin{array}{cccc}
G^R_{11} & F^R_{11} & G^R_{12} & F^R_{12} \\
F^R_{11} & -G^R_{11} & F_{12} & -G_{12}^R \\
G^R_{21} & F^R_{21} & G_{22}^R & F_{22}^R \\
F^R_{21} & -G^R_{21} & F_{22}^R & -G_{22}^R \\
\end{array}\right).
\label{mtrx}
\end{eqnarray}
In addition, each of the matrix elements in Eq. (\ref{mtrx}) is itself a matrix in the channel space. For instance,
$G^R_{11}$ and $G^R_{12}$ are respectively $N_1\times N_1$ and $N_1\times N_2$ matrices, while
matrices $G^R_{21}$ and $G^R_{22}$ have dimensions $N_2\times N_1$ and $N_2\times N_2$ respectively.

As the Keldysh Green function (\ref{GS}) depends neither on time nor on the quasiparticle distribution function, it commutes with the phase factors $e^{\pm i\varphi_r(t)}$
entering the self-energies. This observation combined with the multiplication rule for the $Q-$matrices,  $\hat Q_1\hat Q_2=\hat 1-\hat Q_1+\hat Q_2$,
allows us to express the operator $\check K$ (\ref{K1}) as a linear combination of these matrices:
\begin{eqnarray}
&& \check K(t,t') = [(K_R+ K_A)/2]\otimes \hat 1\,\delta(t-t')
\label{K18}\\ &&
-\, i(\Gamma_1/2) K_R X_1 K_A
\otimes\, e^{i\varphi_1(t)}\hat\sigma_z\hat Q_1(t-t')\hat\sigma_ze^{-i\varphi_1(t')}
\nonumber\\ &&
-\, i(\Gamma_1/2)  K_R X_2 K_A
\otimes\, e^{-i\varphi_1(t)}\hat\sigma_z\hat Q_1(t-t')\hat\sigma_ze^{i\varphi_1(t')}
\nonumber\\ &&
-\, i(\Gamma_2/2) K_R X_3 K_A
\otimes\, e^{i\varphi_2(t)}\hat\sigma_z\hat Q_2(t-t')\hat\sigma_ze^{-i\varphi_2(t')}
\nonumber\\ &&
-\, i(\Gamma_2/2) K_R X_4 K_A
\otimes\, e^{-i\varphi_2(t)}\hat\sigma_z\hat Q_2(t-t')\hat\sigma_ze^{-i\varphi_2(t')}.
\nonumber
\end{eqnarray}
Here $4\times 4$ matrices $X_j$ are defined as follows
\begin{eqnarray}
X_1 = \left(\begin{array}{cccc}
G^R_{11} & F^R_{11} & G^R_{12} & F^R_{12} \\
0 & 0 & 0 & 0 \\
0 & 0 & 0 & 0 \\
0 & 0 & 0 & 0 \\
\end{array}\right),
\nonumber
\end{eqnarray}
\begin{eqnarray}
X_2 = \left(\begin{array}{cccc}
0 & 0 & 0 & 0 \\
F^R_{11} & -G^R_{11} & F^R_{12} & -G^R_{12} \\
0 & 0 & 0 & 0 \\
0 & 0 & 0 & 0 \\
\end{array}\right),
\nonumber
\end{eqnarray}
\begin{eqnarray}
X_3 = \left(\begin{array}{cccc}
0 & 0 & 0 & 0 \\
0 & 0 & 0 & 0 \\
G^R_{21} & F^R_{21} & G^R_{22} & F^R_{22} \\
0 & 0 & 0 & 0 \\
\end{array}\right),
\nonumber
\end{eqnarray}
\begin{eqnarray}
X_4 = \left(\begin{array}{cccc}
0 & 0 & 0 & 0 \\
0 & 0 & 0 & 0 \\
0 & 0 & 0 & 0 \\
F^R_{21} & -G^R_{21} & F^R_{22} & -G^R_{22} \\
\end{array}\right),
\nonumber
\end{eqnarray}
while $4\times 4$  matrices $K_{R,A}$ read
\begin{eqnarray}
K_{R}=K_A^*=\left[ 1 +  i\frac{\Gamma_1}{2}(X_1+X_2)+ i\frac{\Gamma_2}{2}(X_3+X_4) \right]^{-1}.
\label{KR}
\end{eqnarray}

Having established the expression for the operator $\check K$ we
can now substitute it into the action (\ref{S2}). Proceeding along these lines and going trough a straightforward but rather tedious calculation we arrive at the result which still turns out to be too complicated for our present purposes. Further simplification amounts to neglecting local interference terms involving the products of the Green functions $G_{11}^R$, $F_{11}^R$, $G_{22}^R$ and $F_{22}^R$. Technically this step is equivalent to replacing these Green functions by their averaged-over-disorder values, which read
\begin{eqnarray}
\langle G_{11}^R\rangle=\langle G_{22}^R\rangle=0,\;\;\; \langle F_{11}^R\rangle=\langle F_{22}^R\rangle=-\pi\nu_S.
\end{eqnarray}
After this step we immediately arrive at the central result of this section
\begin{eqnarray}
iS_T &=& iS_{11}+iS_{22}+iS_{12},
\label{STfinal}
\end{eqnarray}
where
\begin{eqnarray}
 iS_{11} = -i\frac{G_{11}}{e^2}\int dt \dot\varphi_1\varphi_1^-
-\int dtdt'\frac{\varphi^-_1(t)\,\tilde{\cal S}_{11}^{tt'}\,\varphi^-_1(t')}{2e^2},
\label{S11}
\end{eqnarray}
\begin{eqnarray}
iS_{12} &=& i\frac{G_{12}}{e^2}\int dt \,\big( \dot\varphi_1 \varphi_2^- + \dot\varphi_2 \varphi_1^- \big)
\nonumber\\ &&
-\,\int dtdt'\frac{\varphi^-_1(t)\,\tilde{\cal S}_{12}^{tt'}\,\varphi^-_2(t')}{e^2}
\label{S12}
\end{eqnarray}
and the term $iS_{22}$ is obtained by interchanging the indices $1\leftrightarrow 2$ in Eq. (\ref{S11}). The functions $\tilde{\cal S}_{rl}^{tt'}$ read
\begin{eqnarray}
\tilde{\cal S}_{11}^{tt'} &=&
G_{11}M(t-t')\big( 1-\beta_1 +\beta_1\cos[2\varphi_1^{tt'}] \big)
\nonumber\\ &&
+\, 2G_{12}M(t-t')\big( \alpha_1 - \eta_1\cos[2\varphi_1^{tt'}] \big)
\nonumber\\ &&
+(G_{12}/2)M(t-t')\big( \kappa_1^+\cos[\varphi_1^{tt'}+\varphi_2^{tt'}]
\nonumber\\ &&
+\, \kappa_1^- \cos[\varphi_1^{tt'}-\varphi_2^{tt'}] \big),
\label{xi11}
\\
\tilde{\cal S}_{12}^{tt'} &=&
-G_{12}M(t-t')\big( 1-\beta_1 +\beta_1\cos[2\varphi_1^{tt'}] \big)
\nonumber\\ &&
-G_{12}M(t-t')\big( 1-\beta_2 +\beta_2\cos[2\varphi_2^{tt'}] \big)
\nonumber\\ &&
+\,(G_{12}/2)M(t-t')\big( \gamma_+\cos[\varphi_1^{tt'}+\varphi_2^{tt'}]
\nonumber\\ &&
-\, \gamma_- \cos[\varphi_1^{tt'} - \varphi_2^{tt'}] \big).
\label{xi12}
\end{eqnarray}
Here we denoted $\varphi_r^{tt'}=\varphi_r(t)-\varphi_r(t')$ and introduced Andreev Fano factors $\beta_r=1-\tau_r$.

The zero bias non-local subgap conductance $G_{12}$ as well as
the parameters $\alpha_r,\eta_r,\kappa_r,\gamma$ are expressed as traces of certain combinations of the matrices $K_R$, $K_A$ and $X_j$, as described in Appendix. In order to reduce them to a tractable form we further assume that normal state resistance $R_{\xi}$ of the superconducting
wire segment of length equal to the superconducting coherence length $\xi$ remains much smaller than normal resistances of the barriers \cite{GKZ}, i.e.
\begin{equation}
e^2N_rT_rR_{\xi}/\pi \ll 1.
\label{condition}
\end{equation}
This condition is usually well satisfied for generic systems. Eq. (\ref{condition}) enables one to treat the Green functions connecting the two junctions as small parameters
and expand the traces in Eqs. (\ref{G12})-(\ref{gamma+}) in powers of $G_{12}^R,G_{21}^R,F_{12}^R$ and $F_{21}^R$.
Keeping the leading corrections $\propto G_{12}^RG_{21}^R$ and $\propto F_{12}^RF_{21}^R$ and
making use of the fact that $G_{12}^RG_{21}^R$ becomes equal to $F_{12}^RF_{21}^R$ after averaging over disorder, we arrive at explicit expressions for the parameters
\begin{eqnarray}
\alpha_r&=&\tau_r(1-2\tau_r)/\sqrt{\tau_1\tau_2},
\nonumber\\
\eta_r&=&2\tau_r(1-\tau_r)/\sqrt{\tau_1\tau_2},
\nonumber\\
\kappa_r^{\pm} &=& \pm (4\tau_r-3) + 1/{\sqrt{\tau_1\tau_2}}\;\; (r=1,2),
\nonumber\\
\gamma_{\pm} &=& \pm 1 +\big({1-2\tau_1-2\tau_2+4\tau_1\tau_2}\big)/{\sqrt{\tau_1\tau_2}},
\label{kg}
\end{eqnarray}
and for zero bias non-local conductance
\begin{eqnarray}
G_{12}=\frac{2e^2N_1N_2\tau_1\tau_2}{\pi^3\nu_S^2} G^R_{12}G^R_{21}.
\label{g12}
\end{eqnarray}
In the case of diffusive superconductors one has to further average Eq. (\ref{g12}) over disorder.  In a simple
quasi-one-dimensional geometry this averaging yields \cite{GKZ} $G_{12}=G_{11}G_{22}R_\xi\,e^{-d/\xi}/2$, where $d$ is the distance between two $NS$ barriers.

It is important to emphasize that {\it all} order terms in $t_r$ are fully accounted for in Eqs. (\ref{S11})-(\ref{kg}),
i.e. our action applies for arbitrary transmission values $T_{1,2}$ (or $\tau_{1,2}$) ranging from zero to one and -- similarly to $NS$ systems \cite{GZ09} -- entering
in the combination $T_r^2 / (2-T_r)^2$ representing Andreev transmissions $\tau_r$.

\section{Langevin equations and shot noise}

The quadratic in  $\varphi^-_{1,2}$ action is equivalent to
the Langevin equations \cite{Schmid,AES,GZ92}
\begin{eqnarray}
C_1\dot v_1+(G_{1}^{\rm sh}+G_{11})v_1-G_{12}v_2=G_1^{\rm sh}V_1+\xi^{\rm sh}_1+\xi_1,
\nonumber\\
C_2\dot v_2+(G_{2}^{\rm sh}+G_{22})v_2-G_{12}v_1=G_2^{\rm sh}V_2+\xi^{\rm sh}_2+\xi_2,
\label{langevin}
\end{eqnarray}
which describe the current balance in our system. Here $\xi_r^{\rm sh}$ are stochastic variables with pair correlators
\begin{equation}
\langle\xi_r^{\rm sh}(t)\xi_r^{\rm sh}(t')\rangle=G_r^{\rm sh}M(t-t')
\end{equation}
describing Gaussian current noise in the shunt resistors,
while the variables $\xi_r$ with the correlators
\begin{equation}
\langle\xi_r(t)\xi_l(t')\rangle=\tilde{\cal S}_{rl}^{tt'}
\label{xiS}
\end{equation}
describe shot noise in $NS$ barriers.
Note that we ignore the electro-magnetic coupling between the two junctions, i.e.
mutual capacitance between the normal wires, propagating modes in the superconducting wire \cite{LY}
and similar effects.

Let us first ''turn off'' electron-electron interactions by taking the limit  $1/G_{1,2}^{\rm sh}\to 0$
and, hence, setting $\varphi_r=eV_rt$. Defining the non-local noise spectrum as ${\cal S}_{rl}(\omega) =2\int dt\, \tilde{\cal S}_{rl}^{tt'} \exp (-i\omega (t-t'))$, from Eq. (\ref{xi12})
we obtain
\begin{eqnarray}
&& {\cal S}_{12}(\omega ) =  -2G_{12}(2-\beta_1-\beta_2)W(\omega,0)
\label{noise}\\\ &&
-\,2G_{12}\beta_1W(\omega,2V_1)
- 2G_{12}\beta_2W(\omega,2V_2)
\nonumber\\ &&
+\,G_{12}\gamma_+W(\omega,V_1+V_2)
- G_{12}\gamma_- W(\omega,V_1-V_2),
\nonumber
\end{eqnarray}
where
\begin{eqnarray}
W(\omega,V)=\frac{1}{2}\sum_{\pm}(\omega \pm eV)\coth\frac{\omega\pm eV}{2T}.
\label{coth}
\end{eqnarray}
This is the first key result of the present paper. Eqs. (\ref{noise}),
(\ref{coth}) fully determine non-local shot noise correlations in $NSN$ structures
at subgap energies $T,\omega, eV_r \ll |\Delta |$. In the tunneling limit $T_{1,2}\ll 1$ we have $\gamma_{\pm}=4/T_1T_2\gg \beta_{1,2}\simeq 1$ and Eq. (\ref{noise}) is dominated by the last two terms which further reduce to the result \cite{Pistolesi} in the limit $\omega \to 0$. One of these contributions $\propto \gamma_-$ is due to EC and describes negative shot noise correlations while the other one $\propto \gamma_+$ comes from CAR and accounts for positive cross-correlations of fluctuating currents. Provided one of the voltages, say $V_1$, equals to zero, these EC and CAR terms exactly cancel each other for any $V_2$, i.e. ${\cal S}_{12}(0)$ tends to zero in the tunneling limit.

At higher transmissions $\tau_{1,2}$ the value $\gamma_-$ becomes negative thus implying {\it positive} cross-correlations produced by
direct electron transfer at such values of $\tau_{1,2}$ in addition to CAR. In the limit of fully transparent barriers $\tau_{1,2}=1$ one has $\beta_{1,2}=\gamma_-=0$. Then Eq. (\ref{noise}) yields
\begin{eqnarray}
{\cal S}_{12}(0)= -8TG_{12}
+2eG_{12} (V_1+V_2)\coth\frac{e(V_1+V_2)}{2T}.
\label{noise1}
\end{eqnarray}
At $T \to 0$ only {\it positive} cross-correlations due to CAR
survive whereas no direct electron transfer contribution to shot noise occurs for fully open barriers.
Accordingly, no non-local shot noise is expected in this case at $V_1=-V_2$ and $T \to 0$. We also 
note that noise correlations in clean $NSN$ systems and
in the specific limit $T_1=T_2$, $V_1=V_2$, $T=\omega = 0$
have been recently studied in Ref. \onlinecite{Melin}.
The result of this paper is consistent with our Eq. (\ref{noise}) in the corresponding limit.

For completeness we also provide the expression for the local noise $S_{11}(\omega)$,
which is given by the Fourier transform of Eq. (\ref{xi11}) and reads
\begin{eqnarray}
&& S_{11}(\omega) = 2G_{11}\left[(1-\beta_1)W(\omega,0)+\beta_1 W(\omega,2V_1)\right]
\nonumber\\ &&
+\, 4G_{12}\left[\alpha_1 W(\omega,0)-\eta_1 W(\omega,2V_1)\right]
\nonumber\\ &&
+\, G_{12}\kappa_1^+ W(\omega,V_1+V_2) + G_{12}\kappa_1^- W(\omega,V_1-V_2).
\end{eqnarray}

\section{ Interaction correction to the current}

Now we again ``turn on'' electron-electron interactions and evaluate the current $I_1$ across the first barrier.
Solving Eqs. (\ref{langevin}) perturbatively in $1/g_r \ll 1$, in the lowest non-trivial order in this
parameter we get
\begin{equation}
I_1=G_{11}V_1-G_{12}V_2-\langle\xi_1\rangle.
\end{equation}
Here the average $\langle \xi_1 \rangle$ does not vanish since according to Eqs. (\ref{xi11}), (\ref{xi12}) the noise $\xi_1$
depends on the phases $\varphi_{1,2}$, which, in turn depend on $\xi_{1,2}$ by virtue of Eqs. (\ref{langevin}). Hence, we obtain
\begin{eqnarray}
\langle \xi_1 \rangle =
\left\langle \delta\varphi_1\;{\partial\xi_1}/{\partial\varphi_1}\right\rangle
+ \left\langle \delta\varphi_2\;{\partial\xi_1}/{\partial\varphi_2}\right\rangle,
\label{av1}
\end{eqnarray}
where the phase fluctuations $\delta\varphi_r$ are found from Eqs. (\ref{langevin}) and read
\begin{eqnarray}
\delta\varphi_r(t)=e\int_{-\infty}^{t}dt' \frac{1-e^{-(t-t')/\tau_{RC}}}{G_r^{\rm sh}}\xi_r(t').
\end{eqnarray}
Here we have assumed $G_{12}\ll G_{rr} \ll G_r^{\rm sh}$ and introduced the $RC-$time
$\tau_{RC}=C_r/G_{r}^{\rm sh}$. Substituting this expression into  Eq. (\ref{av1}) we find
\begin{eqnarray}
\frac{\langle \xi_1 \rangle}{e} =\sum_{r=1,2}\int_{-\infty}^{t}dt'
\frac{1-e^{-\frac{t-t'}{\tau_{RC}}}}{G_r^{\rm sh}}
\left.\frac{\partial\langle\xi_1(t)\xi_r(t')\rangle}{\partial\varphi_r(t)}\right|_{\varphi_r=eV_rt}.
\label{av}
\end{eqnarray}
Making use of Eqs. (\ref{xi11}), (\ref{xi12}) and performing the time integral in Eq. (\ref{av}) we get
the current through the first junction in the form
\begin{eqnarray}
&& I_1 = G_{11}V_1-G_{12}V_2
\nonumber\\ &&
-\,\frac{2 G_{11}\beta_1-4G_{12}\eta_1}{g_1}F_0(2V_1)
+ \frac{2G_{12}\beta_2}{g_2}F_0(2V_2)
\nonumber\\ &&
-\delta G_{+}F_0(V_1+V_2) - \delta G_{-}F_0(V_1-V_2),
\label{i1}
\end{eqnarray}
where $\delta G_{\pm}=G_{12}\left({\kappa^{\pm}_1}/{g_1}+{\gamma_{\pm}}/{g_2}\right)$ and
\begin{eqnarray}
&& F_0(V)=\,{\rm Re}\,\left[
- V\Psi\left(1+i\frac{eV}{2\pi T}\right)
\right.
\nonumber\\ &&
\left.
+\,\left( V-\frac{i}{e\tau_{RC}} \right)\Psi\left(1+\frac{1}{2\pi T\tau_{RC}}+i\frac{eV}{2\pi T}\right)
\right].
\end{eqnarray}
Accordingly the local and non-local differential conductances read
\begin{eqnarray}
&& \frac{\partial I_1}{\partial V_1} = G_{11}-\frac{4 G_{11}\beta_1-8G_{12}\eta_1}{g_1}F(2V_1)
\nonumber\\ &&
-\,\delta G_{+}F(V_1+V_2) -\delta G_{-}F(V_1-V_2),
\label{di1dv1}
\end{eqnarray}
and
\begin{eqnarray}
&& \frac{\partial I_1}{\partial V_2} = -G_{12}\left[1-\frac{4\beta_2}{g_2}F(2V_2)\right]
\nonumber\\ &&
-\,\delta G_{+}F(V_1+V_2) +\delta G_{-}F(V_1-V_2),
\label{di1dv2}
\end{eqnarray}
where
\begin{eqnarray}
&& F(V) =\,{\rm Re}\,\left[ \Psi\left(1+\frac{1}{2\pi T\tau_{RC}}+i\frac{eV}{2\pi T}\right)
\right.
\nonumber\\ &&
+\, \left(\frac{1}{2\pi T\tau_{RC}}+i\frac{eV}{2\pi T}\right)\Psi'\left(1+\frac{1}{2\pi T\tau_{RC}}+i\frac{eV}{2\pi T}\right)
\nonumber\\ &&
\left.
-\, \Psi\left(1+i\frac{eV}{2\pi T}\right)- i\frac{eV}{2\pi T}\Psi'\left(1+i\frac{eV}{2\pi T}\right)\right],
\label{psi}
\end{eqnarray}
and $\Psi(x)$ is the digamma function. Eqs. (\ref{di1dv1}), (\ref{di1dv2}) together with Eq. (\ref{psi}) define the conductance matrix of our $NSN$ device and represent the second key result of our paper.

In the interaction correction to the local conductance in  (\ref{di1dv1}) we recover the Coulomb blockade term \cite{GZ09} $\propto \beta_1$ and, in addition, three non-local contributions. The first of them $\propto \eta_1$ enhances the conductance, while the second one $\propto \delta G_{+}$ provides additional Coulomb suppression of $\partial I_1/\partial V_1$. The last term $ \propto \delta G_{-}$ can be both positive (at $\tau_{1,2}\ll 1$) and negative (at bigger $\tau_{1,2}$) implying the tendency to Coulomb anti-blockade in the latter case. The first term $\propto \beta_2$ in Eq. (\ref{di1dv2}) has an opposite sign as compared to $G_{12}$ (thus implying Coulomb blockade), while the second one $\propto \delta G_+$ yields Coulomb anti-blockade. Finally, the third
$\propto \delta G_-$ tends to suppress or enhance the absolute value of the non-local conductance respectively for $\delta G_->0$ and $\delta G_-<0$.

\begin{figure}
\includegraphics[width=9cm]{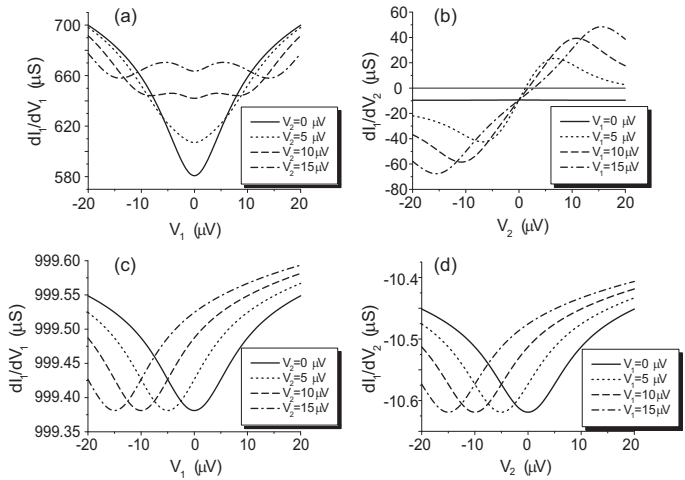}
\caption{Local (a,c) and non-local (b,d) differential conductances defined respectively in Eqs. (\ref{di1dv1}) and (\ref{di1dv2}).
The parameters of the system are: $T=20$ mK, $G_{11}=1$ mS, $G_{12}=10$ $\mu$S, $g_1=g_2=516$.
The dependencies (a) and (b) are plotted at small transmissions $T_1=0.063$, $T_2=0.11$, while the graphs (c) and (d) correspond to fully open barriers $T_1=T_2=1$. }
\label{Gloc}
\end{figure}

The origin of each of the terms in both interaction corrections can easily be identified from the corresponding shot noise correlators (\ref{xi11}), (\ref{xi12}) and (\ref{noise}) illustrating again a fundamental relation between shot noise and Coulomb effects in electron transport. This relation turns out to be considerably more
complicated than in the local case. In the tunneling limit
$T_{1,2} \ll 1$ in Eq. (\ref{di1dv1}) the non-local terms add up to the local one and $\partial I_1/\partial V_1$ evolves from a typical Coulomb blockade V-like dependence at small $V_2$ towards a new W-like one (with extra minima at $V_1=\pm V_2$) at higher $V_2$ (Fig. \ref{Gloc}a). In Eq. (\ref{di1dv2}), in contrast, the last two terms exactly cancel each other for $V_1 \to 0$ and any $V_2$ since $\delta G_{+}=\delta G_{-}$. This cancellation has the same origin as that of EC and CAR contributions to shot noise discussed above. For nonzero $V_1$ and $V_2$ the last two terms in Eq. (\ref{di1dv2}) do not cancel anymore and the curve $\partial I_1/\partial V_2$ approaches the S-like shape with maximum at $V_1=V_2$ and minimum at $V_1=-V_2$ (Fig. \ref{Gloc}b). In this case the interaction term $\propto \delta G_{\pm}\propto T_1T_2\left(1/{g_1}+1/{g_2}\right)$ can exceed  $G_{12}\propto T_1^2T_2^2$ and, hence, $\partial I_1/\partial V_2$
can change its sign. For fully open contacts with $T_{1,2}=1$  we get $\beta_{1,2}=0$ and $\delta G_-=0$, i.e. only CAR terms containing $\delta G_+=2G_{12}\left(1/{g_1}+1/{g_2}\right)$ survive in Eqs. (\ref{di1dv1}) and (\ref{di1dv2}) implying Coulomb blockade for local conductance $\partial I_1/\partial V_1$
and anti-blockade for non-local conductance $|\partial I_1/\partial V_2|$ in this limit (Figs. \ref{Gloc}c,d).

Finally, we would like to note that in some cases non-linearities in both local and non-local differential conductances
caused by Coulomb interaction may be combined with the zero bias anomalies resulting from the proximity-enhanced electron interference in diffusive normal leads \cite{VZK,HN,Z,GKZ}. 
In this paper we disregarded this effect for the sake of simplicity. In practice it implies that here we considered
the system with weakly disordered or sufficiently thick normal leads and sufficiently resistive barriers. 
If needed, zero-bias anomaly effects \cite{VZK,HN,Z,GKZ} can be included into our analysis in a straightforward manner.

\section{Summary}

In this paper we developed a theory elucidating a non-trivial physical relation between shot noise and Coulomb effects in non-local electron transport in $NSN$ structures. 

We evaluated non-local current-current correlators in such systems
at arbitrary transmissions of $NS$ interfaces and arbitrary frequencies, Eqs. (\ref{noise}), (\ref{coth}). This result
demonstrates that positive cross-correlations in shot noise increase with increasing interface transmissions
and dominate the result for fully open barriers in which case only CAR contribution survives. Positive noise cross-correlations in $NSN$ structures have been convincingly demonstrated in recent experiments \cite{Chandrasekhar}, while no or weak negative cross-correlations have been observed. This picture would qualitatively 
correspond to the case of highly transmitting interfaces,  cf. Eq.  (\ref{noise1}). Note, however, that interface
transmissions in experiments \cite{Chandrasekhar} are reported to be rather small, in which case one would
expect EC-induced negative cross-correlations to dominate the result at $V_1=-V_2$. 

Turning to the effect of electron-electron interactions on non-local electron transport we would like to emphasize
several important new features demonstrated within our analysis. One of them is that in the tunneling limit almost no
effect of Coulomb interaction on non-local conductance is expected if one of the applied voltages, $V_1$ or $V_2$, equals to zero. This effect is directly related to the cancellation between EC and CAR contributions to shot noise in the corresponding limit. For nonzero $V_1$ and $V_2$  no such cancellation exists anymore and the non-local conductance $\partial I_1/\partial V_2$ approaches the S-like shape being enhanced at $V_1\approx V_2$ and partially suppressed at $V_1\approx -V_2$, see Fig. \ref{Gloc}b. Both these features have a clear physical interpretation. Indeed, at $V_1\approx -V_2$ negative cross-correlations due to EC dominate non-local shot noise leading to Coulomb blockade of non-local conductance while at $V_1\approx V_2$ positive cross-correlations due to CAR prevail and Coulomb anti-blockade of non-local transport is observed. At higher interface transmissions only Coulomb anti-blockade of non-local conductance remains (Fig. \ref{Gloc}d), which is again related to CAR-induced positive cross-correlations in shot noise.
 
It is interesting to point out that S-like shaped non-local signal predicted here was indeed observed in experiments \cite{Beckmann2,newexp}. A good agreement between our theory and the results \cite{newexp} argues in favor of electron-electron interactions as a physical reason for the observed feature. Some of the features similar to those predicted here have also been observed in experiments \cite{Chandrasekhar}. More experiments on both non-local
shot noise and non-local electron transport would be desirable in order to quantitatively verify our predictions. 

\vspace{0.5cm}

\centerline{\bf Acknowledgments}

\vspace{0.5cm}

We are indebted to D. Beckmann for making us aware of the results \cite{newexp} prior to publication. This work was supported in part by DFG and by RFBR grant 09-02-00886.

\appendix

\section{Parameters of the action}

For the sake of completeness, let us present general expressions for non-local conductance $G_{12}$ and the parameters $\kappa_r^{\pm}$ and $\gamma_{\pm}$ entering in the effective action (\ref{xi11}), (\ref{xi12}). We have
\begin{eqnarray}
G_{12}=\frac{e^2}{2\pi}\,{\rm tr}\,\big[(X_3-X_4)K_R(X_1-X_2)K_A\big],
\label{G12}
\end{eqnarray}
\begin{eqnarray}
\kappa_1^+&=&(e^2\Gamma_1\Gamma_2/\pi G_{12})\,{\rm tr}\,\big[
2K_RX_4K_AX_1
\nonumber\\ &&
+\,2\Gamma_1^2 K_RX_1K_A(X_1-X_2)K_RX_4K_A(X_1-X_2)
\nonumber\\ &&
-\,2i\Gamma_1(X_1-X_2)K_RX_1K_RX_4K_A
\nonumber\\ &&
+\,2i\Gamma_1X_1K_A(X_1-X_2)K_RX_4K_A
\big],
\label{k+}
\end{eqnarray}
\begin{eqnarray}
\gamma_+&=&({e^2\Gamma_1\Gamma_2}/{\pi G_{12}})\,{\rm tr}\,\big[
K_RX_4K_RX_1+K_AX_4K_AX_1
\nonumber\\ &&
-\,\Gamma_1\Gamma_2K_RX_1K_A(X_1-X_2)K_RX_4K_A(X_4-X_3)
\nonumber\\ &&
-\,\Gamma_1\Gamma_2K_RX_1K_A(X_4-X_3)K_RX_4K_A(X_1-X_2)
\nonumber\\ &&
+\,i\Gamma_2(X_4-X_3)K_RX_1K_RX_4K_A
\nonumber\\ &&
+\,i\Gamma_1(X_1-X_2)K_RX_4K_RX_1K_A
\nonumber\\ &&
-\,i\Gamma_2 X_1K_A(X_4-X_3)K_RX_4K_A
\nonumber\\ &&
-\,i\Gamma_1 X_4K_A(X_1-X_2)K_RX_1K_A
\big].
\label{gamma+}
\end{eqnarray}
The parameters $\kappa_1^-$ and $\gamma_-$ are defined by Eqs. (\ref{k+}) and (\ref{gamma+})
with interchanged matrices $X_3\leftrightarrow X_4$.
The remaining parameters $\alpha_r$ and $\eta_r$ in Eqs. (\ref{xi11}), (\ref{xi12}) are defined in a similar manner. These parameters are less important for our consideration and we omit the corresponding expressions for the sake of brevity.

\end{document}